\documentstyle[11pt,newpasp,twoside]{article}
\index{instructions}
\index{guidelines}
\index{pulsar wind nebulae!G11.2--0.3}
\index{pulsar wind nebulae!G292.0+1.8}
\index{supernova remnants!B0540--69}
\index{pulsar wind nebulae!G184.6--5.8}
\index{pulsar wind nebulae!G130.7+3.1}
\index{pulsar wind nebulae!G29.7--0.3}
\index{pulsar wind nebulae!G320.4--1.2}
\index{pulsar wind nebulae!G54.1+0.3}
\index{pulsar wind nebulae!interactions with supernova remnants}

\newcommand{\Msun}{M_\odot}
\newcommand{\kms}{\rm ~km~s^{-1}}
\newcommand{\ml}{~\Msun ~\rm yr^{-1}}

\def\edcomment#1{\iffalse\marginpar{\raggedright\sl#1\/}\else\relax\fi}
\reversemarginpar

\begin{document}
\title{Pulsar Wind Nebulae and Their Supernovae}
 \author{Roger A. Chevalier}
\affil{Department of Astronomy, University of Virginia, P.O. Box 3818,
Charlottesville, VA 22903, USA}

\begin{abstract}
Young supernova remnants that contain pulsar wind nebulae
provide diagnostics for both the inner part of the supernova
and the interaction with the surrounding medium, providing
an opportunity to relate these objects to supernova types.
Among observed young nebulae, there is evidence for a range
of supernova types, including Type IIP (Crab Nebula and SN 1054)
and Type IIb/IIn/IIL (G292.0+1.8).
\end{abstract}

\section{Introduction}
Recent X-ray satellites
({\it ASCA}, {\it Chandra}, and {\it Newton XMM}) have
substantially increased the  number of observed young PWNe (pulsar wind
nebulae), as well as adding to our knowledge of the nebulae
and their surrounding supernova remnants.
Young PWNe are expected to be interacting with freely
expanding ejecta in the interior of a supernova, providing
a probe of the inner parts of the supernova.
The surrounding supernova remnant is typically interacting
with circumstellar mass loss from the progenitor star.
The combination of this information can be related to our
expections for core-collapse supernovae and their surroundings.
Here, I examine the properties of 8 young PWNe in this context.

\section{Supernovae and their surroundings}

Recent observations of core-collapse supernovae at X-ray
and radio wavelengths have yielded a fairly complete picture
of the circumstellar surroundings of the various supernova
types (Chevalier 2003a).
Although the numbers are still small, Type IIP (plateau) supernovae have
been found to be relatively weak X-ray and radio sources,
with a mass loss density corresponding to a mass loss rate
$\dot M\approx 10^{-6}\ml$ for a wind velocity of $v_w=10\kms$;
the light curves of these supernovae indicate red supergiant progenitors.
The low mass loss is consistent with the fact that the
supernova properties indicate that most of the stellar H envelope
is present at the time of the explosion.
A consequence of this is that the inner heavy element core is
decelerated by the envelope and mixes with it, giving rise
to low velocity H in the supernova.
There is information on the progenitor stars of 2 Type IIP
supernovae, leading to upper mass limits of the progenitors
of $12\Msun$ and $15\Msun$ (Smartt et al. 2003).
Thus, SNe IIP appear to be near the lower mass limit for
core-collapse supernovae.

SN 1987A is another supernova that exploded with a massive
H envelope that was able to decelerate the heavy element ejecta.
However, the progenitor in this case was a blue supergiant
with a mass $\sim 19\Msun$.
The immediate surrounding of the blue supergiant star was determined
by the fast wind from that star but, further out,
beyond the ring at 0.2 pc, there is
evidence for a slow red supergiant wind (Crotts 2000).

Supernovae that show strong interaction with a dense red
supergiant wind are of Types IIn (narrow line), IIL (linear), and IIb
(evolve to Ib/c spectrum);
these types are not mutually exclusive as they refer to different
aspects of the supernova.
In these cases, the presupernova mass loss can go up to
$\dot M\approx 10^{-4}-10^{-3}\ml$ for  $v_w=10\kms$.
In SNe IIb and at least some IIn (e.g., Fransson et al. 2002),
most of the H envelope is lost before the supernova.
The result is that there is no slow H in the supernova;
e.g., in the IIb SN 1993J, H was in the range
$8500-10,000\kms$ (Houck \& Fransson 1996).

The dense wind from a red supergiant progenitor star can extend
out some distance around the progenitor.
The wind comes into pressure equilibrium with the surrounding medium
at a radius (Chevalier \& Emmering 1989)
\begin{equation}
r_{sh}=5.0\left(\dot M\over 5\times 10^{-6}\ml\right)^{1/2}
\left(v_w\over 15\kms\right)^{1/2}
\left(p_i\over 10^3 {\rm cm^{-3} ~K}\right)^{-1/2}  {\rm~pc,}
\end{equation}
where $p_i$ is the external pressure.
This expression assumes that the wind has been steady for at
least a time $r_{sh}/v_w$, or $3.3\times 10^5$ years for the
reference values.
The red supergiant wind may end in a shell, which may have been
observed around SN 1987A (Chevalier \& Emmering 1989).
Beyond the red supergiant wind, there may be a wind bubble created
by the progenitor star when it was on the main sequence.
Although the wind bubble in a low density uniform medium can be
large, interstellar clouds could be relatively close to the progenitor.

The remaining types of core collapse supernovae are the Type Ib  and Ic
explosions, which are thought to have massive star progenitors that
lost their H envelopes.
In this case, no slow H is expected in the supernova, and the immediate
surroundings of the supernova are determined by the 
low density, fast wind typical
of a Wolf-Rayet star ($\dot M\approx 10^{-5}\ml$ and  $v_w\approx 10^3\kms$).
The combination of  stellar mass loss estimates with the fact that
stars with an initial mass $\ga 25\Msun$ collapse to black holes suggests
that Type Ib  and Ic are the result of binary evolution (Heger et al. 2003).
In this case, a substantial part of the H envelope mass loss is due 
to the companion star.

Dahl\'en \& Fransson (1999) have summarized recent work on supernova rates, 
finding the
following intrinsic fractions of core-collapse supernova types: 
SN IIP (0.3), SN IIL (0.30), SN IIn (0.02), SN 1987A-like (0.15), 
SN Ib/c (0.23); the rate of SN 1987A-like events is especially uncertain.
Unless there is a strong correlation of pulsars with supernova types,
PWNe should appear in the young remnants of a variety of supernova types.

\section{Young pulsar wind nebulae}

Table 1 lists 8 young PWNe that are plausibly
interacting with freely expanding ejecta from the supernova
(see Chevalier 2003b for references to these objects).
Tentative supernova identifications have been given in 4 cases,
but only the identification of the Crab Nebula with SN 1054 can
be considered secure.
In two other cases (0540--69 and G292.0+1.8), there is an age
estimate from the expansion of optical filaments.
$R_{snr}$ is the radius of the surrounding supernova remnant
except where no surrounding remnant is observed, in which case
the radius of the PWN is given (in parentheses).
$V_{snr}$ is the average velocity of the remnant obtained by
dividing $R_{snr}$ by the age.

\begin{table}

\caption{Properties of Young Pulsar Nebulae }

\begin{tabular}{cccccc}
 \tableline
 Supernova &  $P/2\dot P$ &  Age   & SN &  $R_{snr}$ & $V_{snr}$  \\
  Remnant  &  (yr)  &  (yr)  &       &  (pc)  & ${\rm km~s^{-1}} $   \\
 \tableline
  0540--69 &  1660 &      760  &  & 9  & 11,600   \\
  3C 58 (G130.7+3.1)    &  5390   &    821  & 1181  & (3.3)   & (3,900)    \\
 Crab (G184.6--5.8) &   1240  &    948   &  1054  &  (2)    & (2,100)   \\
  Kes 75 (G29.7--0.3)      &   723  &  &    &  9.7  & 13,000 \\
  G11.2--0.3   & 2890  & 1616 & 386  &  3.3  & 2,000  \\
  G292.0+1.8  & 24,000 & $\la 1700$  &   &  5.8  & 3,200 \\
  MSH 15-52 (G320.4--1.2) & 1700  &  1817  & 185  & 20  & 10,800  \\
  G54.1+0.3    & 2890 &  &   & (1.3) & (440)  \\
 \tableline
 \tableline

\end{tabular}

\end{table}

There are several arguments for associating the Crab Nebula and SN 1054
with a SN IIP.
The presence of H in the ejecta suggests that the core material
was decelerated by the H envelope, as expected in a SN IIP.
The abundances in the Crab imply a progenitor mass of $\sim 9\Msun$
(Nomoto 1985), which is compatible with SN IIP.
Finally, the lack of a shell around the Crab implies that the supernova is
interacting with a low density surrounding medium.
The low radio and X-ray luminosities of SN IIP suggest that they
have a relatively low density wind, which may not extend far from
the progenitor (eqn. [1]).
The supernova may have swept through the wind and is now in a low
density bubble.
This scenario for SN 1054 was outlined in Chevalier (1977), but the
absence of any observable interaction around the Crab has frustrated
attempts at verification.

There are no ejecta observations in 3C58, but the presence of
slow moving, N-rich flocculi (Fesen, Kirshner, \& Becker 1987) 
is suggestive of a clumpy red
supergiant wind surrounding the progenitor.
A SN IIP explosion is thus a possibility, with the lack of
a surrounding shell being due to a relatively small amount of
mass loss.
For G54.1+0.3, the lack of a surrounding shell suggests that it
is not a SN IIb/IIn/IIL, but there is not enough information on the
object to be more definite.

The pulsar and PWN in 0540--69 have many similarities to the Crab
and its pulsar, but there is a shell present in this case.
The presence of H in the shell swept up by the PWN has been
controversial (Kirshner et al. 1989); if it is present, a SN IIP
explosion is indicated.
The progenitor mass would be somewhat larger that for SN 1054
because of the enhanced O emission from the ejecta ($\ga 13\Msun$
initial mass).
The large size of the shell and the high average velocity imply
that the supernova encountered little circumstellar matter before
running into denser, inhomogeneous surrounding gas; i.e.
the supernova shock traversed a stellar wind bubble.  
The large sizes and velocities observed in MSH 15-52 and Kes 75
suggest a similar situation in these objects, implying that they
did not result from Type IIb/IIn/IIL supernovae.
On the basis of the high energy/mass ratio deduced for MSH 15-52,
Gaensler et al. (1999) suggested a SN Ib origin; the finding of
H-free ejecta would help to confirm this designation.

The remaining two objects, G11.2--0.3 and G292.0+1.8, have
remnants with both a small radius and low velocity, suggesting
strong circumstellar interaction and a SN IIb/IIn/IIL origin.
The red supergiant wind from the progenitor star can extend to
the observed radii, so this is the likely ambient medium.
Strong mass loss is consistent with the absence of Balmer lines
in the spectrum of G292.0+1.8 (Goss et al. 1979), implying that
most of the H envelope had been lost before the supernova.

These considerations show that the supernova remnants around 
young PWNe are likely to be strongly influenced by the presupernova
mass loss.
Future observations of ejecta abundances and additional observations of
the circumstellar interaction should allow clearer associations
with supernova types.

\acknowledgments
This work was supported in part by NASA grant NAG5-13272.


\begin{references}

\reference Chevalier, R.~A.\ 1977, in Supernovae, ed. D. N. Schramm
(Dordrecht: Reidel), 53
\reference Chevalier, R.~A.\ 2003a, in From 
Twilight to Highlight: The Physics of Supernovae, ed.
W. Hillebrandt \& B. Leibundgut (Berlin: Springer), 299 
\reference Chevalier, R.~A.\ 2003b, in High Energy Studies of Supernova
Remnants and Neutron Stars, (COSPAR 2002), Adv. Sp. Res., in press 
(astro-ph/0301370)
\reference Chevalier, R.~A., \& Emmering, E.~T.\ 1989, \apj, 342, L75
\reference Crotts, A.~P.~S.\ 2000, in Asymmetrical Planetary Nebulae II,
ed. J. H. Kastner, N. Soker, \& S. A. Rappaport (San Francisco: ASP ), 445
\reference Dahl{\' e}n, 
T.~\& Fransson, C.\ 1999, \aap, 350, 349 
\reference Fesen, R. A., Kirshner, R. P., \& Becker, R. H.\ 1987, in Supernova
Remnants and the Interstellar Medium, ed. R. S. Roger \& T. L. Landecker
(Cambridge: CUP), 55
\reference Fransson, C., et al.\ 2002, \apj, 572, 350
\apj, 569, 878 
\reference Gaensler, B.~M., 
Brazier, K.~T.~S., Manchester, R.~N., Johnston, S., \& Green, A.~J.\ 1999, 
\mnras, 305, 724 
\reference  Goss, W. M., Shaver, P. A., Zealey, W. J., Murdin, P., \&
Clark, D. H.\ 1979, \mnras, 188, 357
\reference Heger, A., Fryer, C.~L., 
Woosley, S.~E., Langer, N., \& Hartmann, D.~H.\ 2003, \apj, 591, 288 
\reference Houck, J.~C., \& Fransson, C.\ 1996, \apj, 456, 811
\reference Kirshner, R. P., Morse, J. A., Winkler, P. F., \& Blair, W. P.\
1989, \apj, 342, 260
\reference Nomoto, K.\ 1985, in The Crab Nebula and Related Supernova Remnants,
ed. M. C. Kafatos \& R. C. B. Henry, (Cambridge: CUP), 97
\reference Smartt, S.~J., Maund, 
J.~R., Gilmore, G.~F., Tout, C.~A., Kilkenny, D., \& Benetti, S.\ 2003, 
\mnras, 343, 735 


\end{references}
\end{document}